\documentclass[footinbib,a4paper,aps,superscriptaddress,reprint,twocolumn,preprintnumbers,amsmath,amssymb,nobalancelastpage,10pt]{revtex4-2}

\usepackage{amsmath}
\usepackage{verbatim}
\usepackage{graphicx}
\usepackage{wasysym}

\usepackage{color}
\usepackage[dvipsnames]{xcolor}
\usepackage{braket}
\usepackage{yfonts}
\usepackage{soul}
\usepackage[normalem]{ulem}
\usepackage{dsfont}
\usepackage[colorlinks=true,citecolor=blue,linkcolor=blue,urlcolor=blue]{hyperref}

\usepackage{amsmath}
\usepackage{bm}
\usepackage{amsfonts}

\usepackage{graphicx}
\graphicspath{{Figures}}

\begin{document}

\def \IBK{Institute for Theoretical Physics, University of Innsbruck, 6020 Innsbruck, Austria}
\def \IQOQI{Institute for Quantum Optics and Quantum Information of the Austrian Academy of Sciences, 6020 Innsbruck, Austria}

\title{U($1$) lattice gauge theory and string roughening on a triangular Rydberg array}

\author{Lisa Bombieri}\email{lisa.bombieri@uibk.ac.at}\affiliation{\IBK}\affiliation{\IQOQI}
\author{Torsten V. Zache}\affiliation{\IBK}\affiliation{\IQOQI}
\author{Hannes Pichler }\affiliation{\IBK}\affiliation{\IQOQI}
\author{Daniel Gonz\'alez-Cuadra}\email{daniel.gonzalez@ift.csic.es}
\affiliation{Department of Physics, Harvard University, Cambridge, MA 02138, USA}
\affiliation{Instituto de F\'isica Te\'orica UAM-CSIC, C. Nicol\'as Cabrera 13-15, Cantoblanco, 28049 Madrid, Spain}

\begin{abstract}
Lattice gauge theories (LGTs) describe fundamental interactions in particle physics. A central phenomenon in these theories is confinement, which binds quarks and antiquarks into hadrons through the formation of string-like flux tubes of gauge fields. Simulating confinement dynamics is a challenging task, but recent advances in quantum simulation are enabling the exploration of LGTs in regimes beyond the reach of classical computation. For analog devices, a major difficulty is the realization of strong plaquette interactions, which generate string fluctuations that can drive a roughening transition. Understanding string roughening---where strong transversal functions lead to an effective restoration of translational symmetry at long distances---is of central importance in the study of confinement.
In this work, we show that string roughening emerges naturally in an analog Rydberg quantum simulator. We first map a triangular Rydberg array onto a $(2+1)$D U$(1)$ LGT where plaquette terms appear as first-order processes. We study flux strings connecting static charges and demonstrate that, near a deconfined quantum critical point, the string exhibits logarithmic growth of its transverse width as the separation between charges increases, along with the universal L\"uscher correction to the confining potential---both signatures of string roughening. Finally, we investigate the real-time dynamics of an initially rigid string, observing large fluctuations after quenching into the roughening regime, as well as string breaking via particle-pair creation. Our results indicate that rough strings can be realized in experimentally accessible quantum simulators, opening the door to detailed studies of how strong fluctuations influence string-breaking dynamics.

\end{abstract}

\maketitle

\section{Introduction}

The fundamental forces of nature are described by gauge theories, in which local gauge symmetries give rise to force-carrying gauge fields mediating interactions between particles~\cite{Gross_2023}, and they also emerge as effective descriptions of diverse phenomena in condensed matter physics~\cite{Wen_2017, Sachdev_2019}. Lattice gauge theories (LGTs) provide a lattice regularization of these models~\cite{Montvay_1997}, making them amenable to numerical simulations. However, simulating these theories---especially their real-time dynamics in non-perturbative regimes---is extremely challenging for classical computers. This has motivated the development of quantum simulators~\cite{Altman_2021, Daley_2022}, programmable quantum systems that can simulate the dynamics of LGTs in regimes inaccessible to classical approaches~\cite{Banuls_2020, Aidelsburger_2022, Klco_2022, Bauer_2023, Di_Meglio_2024}.

Recent advances in assembling, manipulating, and measuring quantum systems across multiple platforms, such as neutral atoms~\cite{Gross_2017, Kaufman_2021}, trapped ions~\cite{Monroe_2021}, and superconducting qubits~\cite{Kjaergaard_2020}, enable access to novel regimes in quantum simulation.
These platforms have allowed the investigation of LGTs in different regimes~\cite{Martinez_2016, Klco_2018, Schweizer_2019, Kokail_2019, Mil_2020, Yang_2020, Nguyen_2021, Zhou_2021, Frolian_2022, Mildenberger_2025, Meth_2025, Liu_2025, Gonzalez_2025, Cochran_2025, Lerose_2024, Datla_2025, Schuhmacher_2025, Cobos_2025,  Xiang_2025, Mark_2025}, with particular interest in the confinement phenomenon. In quantum chromodynamics (QCD)~\cite{Gross_2023}, this phenomenon underlies the confinement of quarks into composite hadrons due to the formation of string-like flux tubes of gluons. At large separations, the energy stored in the flux string becomes sufficient to break the string by producing additional quark–antiquark pairs. While in high-energy particle colliders these mechanisms are probed  through the detection of jets of secondary particles~\cite{Gross_2023, Bali_2005, Altmann_2025, Florio_2023}, quantum simulators allow the direct exploration of string-breaking dynamics with spatio-temporal resolution~\cite{Lerose_2024, Liu_2025, Gonzalez_2025, Cochran_2025, Cobos_2025, Alexandrou_2025}.

Recently, different experiments have started to address string breaking in (2+1)D LGTs~\cite{Gonzalez_2025, Cochran_2025, Cobos_2025, Crippa_2024}. Beyond the one-dimensional case, simulating LGTs becomes more challenging due to the transverse fluctuations of the flux string~\cite{Zohar_2022}. These fluctuations are the basis of string roughening: while the string connecting a particle–antiparticle pair remains rigid under a strong confining potential, as the potential weakens, transverse fluctuations grow and can effectively restore the continuous translational symmetry and delocalize the string. When this occurs, the string is referred to as `rough'~\cite{Luscher_1981_1, Drouffe_1981, Drouffe_1981_1, Hasenfratz_1981, Munster_1981, Munster_1981_1, Caselle_1996, M_Hasenbusch_1997, Gliozzi_2010, Xu_2025, DiMarcantonio_2025}, characterized by a logarithmically diverging width with increasing particle separation and by a universal correction to the classical linear confining potential, known as L\"uscher term~\cite{Luscher_1981}. These fluctuations, however, are driven by multi-body plaquette interaction terms, which are difficult to implement in current analog quantum simulators~\cite{Dai_2017}. In such systems they typically emerge as weak effective terms generated at high orders in perturbation theory~\cite{Halimeh_2025}, making the observation of string roughening an outstanding challenge.

\begin{figure*}
    \centering
    \includegraphics[width=\textwidth]{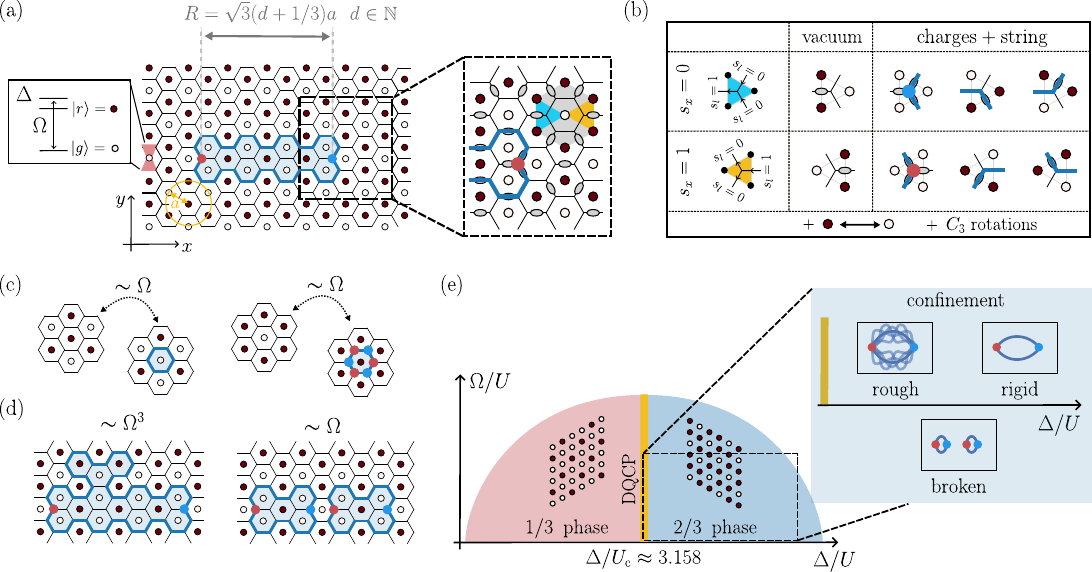}
    \caption{(a) Rydberg atoms trapped in optical tweezers arranged on a triangular geometry with lattice spacing $a$. The figure depicts a $2/3$-ordered configuration of Rydberg excitations, corresponding to the LGT bare vacuum on the dual hexagonal lattice. Atoms in the region marked in light blue have been flipped ($\ket{g}\leftrightarrow \ket{r}$). In the LGT picture, this atomic configuration corresponds to a positive $Q = +2$ (red dot) and a negative $Q = -2$  (blue dot) charge connected by two rigid strings of electric field (dark blue lines) and separated by a distance $R=\sqrt{3}(d+1/3)a\approx5.77a$, corresponding to $d=3$. Inset: charges and strings reside on the sites and links of the dual hexagonal lattice, respectively. Strings form in correspondence of flipped dimers (gray ellipses) relative to the vacuum. The six-site unit cell is highlighted in gray. (b) Equivalence between the atomic configurations and the corresponding gauge-invariant states for two sites of the unit cell, highlighted in blue and orange in (a), respectively. Global spin flips leave the configurations unchanged, and the configurations associated to the remaining four sites are obtained through $C_3$ rotations. (c) First order processes in the Rabi frequency $\Omega$ induce the formation of string loops and charged loops in the bare vacuum. (d) Third order processes in $\Omega$ lead to transversal fluctuations of the string. First order processes in $\Omega$ lead to string breaking. (e) Schematic ground-state phase diagram for the triangular Rydberg array. In the absence of defects (static charges), the system hosts two ordered phases---the $1/3$- and the $2/3$-phase---separated by a DQCP (yellow line). Inset: introducing two static charges (red and blue circle) in the $2/3$ phase generates two strings of electric field (dark blue lines) connecting them. Deep in the phase, the string is `rigid' with a width that remains constant as the distance between the two charges increases. Approaching the DQCP, the string becomes `rough' with its width increasing logarithmically with distance.  Based on the distance the string can break at different values of $\Delta/U$, leading to the creation of pairs of dynamical charges. }
    \label{fig:Fig1_Model}
\end{figure*}

In this work, we propose Rydberg atoms trapped in optical tweezers~\cite{Browaeys_2020} operating in the analog quantum simulator mode~\cite{Chen_2021, Shaw_2024,Manovitz_2025,Anand_2024} as a platform where string roughening emerges naturally, posing it as a promising candidate to explore the interplay between string fluctuations and string breaking. Rydberg quantum simulators have already explored LGTs in both $(1+1)$D~\cite{Bernien_2017, Surace_2020, Datla_2025, Xiang_2025, Mark_2025} and $(2+1)$D~\cite{Semeghini_2021, Gonzalez_2025}. However, these implementations do not include plaquette interactions~\cite{Celi_2020, Homeier_2022, Feldmeier_2024, Cheng_2024, Zhou_2025}, which are crucial for accessing the roughening regime. In contrast, here we consider an array of Rydberg atoms arranged on a triangular lattice and coupled via van der Waals interactions, a setup already available in current experiments~\cite{Scholl_2021}, and we identify a mapping to a U$(1)$ LGT in which the plaquette interactions emerge at first order (rather than sixth order~\cite{Gonzalez_2025}) in perturbation theory. 
Previous studies~\cite{Scholl_2021, Bombieri_2025} have shown that the ground-state phase diagram of this system hosts two ordered phases at $1/3$ and $2/3$ filling, separated by a deconfined quantum critical point (DQCP) [see Fig.~\ref{fig:Fig1_Model}(e)]. Here, we refine the physical interpretation of the notion of `deconfinement' and show that string roughening appears as we approach the DQCP.

The article is organized as follows. In Sec.~\ref{sec:model}, we map the triangular Rydberg array into a U$(1)$ LGT on the dual hexagonal lattice, where we consider the ground state deep into the $2/3$ phase as the vacuum of the theory and show that pairs of charge excitations are confined by flux strings. Deep within this phase, the string remains rigid; however, as the system approaches the DQCP, it begins to fluctuate and eventually becomes rough [Fig.~\ref{fig:Fig1_Model}(e)]. We demonstrate this quantitatively in Sec.~\ref{sec:roughening} using tensor-network methods, showing that the width of the string grows logarithmically with the separation between excitations and that the confining potential acquires a correction to its classical linear form consistent with the universal L\"uscher term---key signatures of a rough string. 
In Sec.~\ref{sec:dynamics}, we investigate the out-of-equilibrium dynamics of an initially rigid string following a quench, observing strong string fluctuation in the roughening region, as well as string breaking via particle-pair creation. We summarize our results in Sec.~\ref{sec:conclusions}, where we comment on the implications for current quantum simulation experiments.

\section{Model and LGT mapping}\label{sec:model}

In this section, we detail the mapping between Rydberg atoms on a triangular lattice and a U$(1)$ LGT on the dual hexagonal lattice. 
Excitations in the ordered phases of the atomic model are mapped to configurations of charges connected by flux strings in the LGT.
Furthermore, we show that charges are confined within the ordered phases in the classical limit of zero Rabi frequency and that quantum fluctuations can induce both string breaking and string fluctuations.

\subsection{Rydberg atoms on a triangular array} 
We consider neutral atoms trapped in optical tweezers arranged on a triangular lattice with lattice constant $a$ [Fig.~\ref{fig:Fig1_Model}(a)]. Two electronic states---the ground state $\ket{g}$ and a highly excited Rydberg level $\ket{r}$---are laser coupled, giving rise to many-body dynamics described by the Hamiltonian
\begin{equation}
    \hat{H}_{\rm Ryd}= \frac{\Omega}{2} \sum_{{\bf j}} \hat{\sigma}^{x}_{{\bf j}} -\Delta \sum_{{\bf j}} \hat{n}_{{\bf j}} + \sum_{{\bf i} > {\bf j}} U_{{\bf i} {\bf j}}\, \hat{n}_{{\bf i}} \hat{n}_{{\bf j}},
    \label{eq:hamiltonain_rydberg}
\end{equation}
where $\hat{n}_{{\bf j}}=\ket{r}_{{\bf j}}\bra{r}$, $\hat{\sigma}^{x}_{{\bf j}}=\ket{r}_{{\bf j}}\bra{g}+\ket{g}_{{\bf j}}\bra{r}$, and $U_{{\bf i} {\bf j}}=C_6/|x_{\bf i}-x_{\bf j}|^6$ is the van der Waals interaction strength between excited atoms at positions $x_{\bf i}$ and $x_{\bf j}$. 
As shown in previous works, this model features a rich ground-state phase diagram, with ordered phases at $1/3$ and $2/3$ Rydberg filling~\cite{Scholl_2021,Guo_2023}---hereafter named $1/3$ and $2/3$ phase. In quasi-1D geometries, these two phases are separated by a DQCP~\cite{Bombieri_2025}, located around $\Delta/U_{\rm c}\approx3.158$ (taking into account interactions up to the third-nearest-neighbor), where $U$ is the strength of the nearest-neighbor interaction [Fig.~\ref{fig:Fig1_Model}(e)].

\subsection{Mapping to a U$(1)$ LGT}
In this work, we investigate the confinement of low-energy excitations in the $2/3$ phase. To this end, we first map atomic configurations onto dimer variables defined on the dual hexagonal lattice, which are then further mapped to configurations of a U$(1)$ LGT. Within this framework, defects on top of the ordered atomic configuration map to pairs of charges connected by electric flux lines (strings).

\subsubsection{Mapping to dimer configurations} 

Similarly to the Ising model on a triangular lattice~\cite{Moessner_2001, Moessner_2001_1},
configurations of Rydberg atoms can be mapped onto dimer configurations on the dual hexagonal lattice, up to a global spin flip. We introduce spin variables $\hat{\sigma}^z_{\bf i}=2\hat{n}_{\bf i}-1$, which take the value $-1$ for atoms in the ground state $\ket{g}$ and $+1$ for atoms in the Rydberg state $\ket{r}$. Whenever a link $l=\left({\bf i}, {\bf j}\right)$ connects two neighboring atoms with aligned spins ($\hat{\sigma}^z_{\bf i} \hat{\sigma}^z_{\bf j} = 1$), the corresponding link on the dual lattice is occupied by a dimer; otherwise ($\hat{\sigma}^z_{\bf i} \hat{\sigma}^z_{\bf j} = -1$), it is unoccupied, as illustrated in the inset of Fig.~\ref{fig:Fig1_Model}(a). The associated dimer-occupation operator is
\begin{equation}
    \hat{D}_{l} = \frac{\mathbb{I}+\hat{\sigma}^z_{\bf i} \hat{\sigma}^z_{\bf j}}{2} = \mathbb{I}+2\hat{n}_{\bf i}\hat{n}_{\bf j}-\hat{n}_{\bf i}-\hat{n}_{\bf j},
\end{equation}
where $\mathbb{I}$ is the identity operator.
Each atomic configuration maps to a unique dimer configuration on the dual lattice, with the mapping being two-to-one, since global spin flips leave the dimer configuration unchanged. In particular, both the $1/3$ and $2/3$ phases correspond to the star phase of the dimer model on the hexagonal lattice~\cite{Schlittler_2017}. This phase satisfies the dimer constraint, i.e., each vertex is touched by exactly one dimer, which we interpret as a local U($1$) symmetry in the following.

\subsubsection{Mapping to U$(1)$ LGT configurations} 
A defining feature of a LGT is its invariance under local gauge transformations, which imposes a local conservation law on physical states. For a U$(1)$ LGT, this takes the form of Gauss’s law, $\nabla \cdot \vec{E}-\rho=0$, which relates the divergence of the electric field to the charge density $\rho$ at each lattice site. In the Hamiltonian formulation~\cite{Kogut_1975}, this constraint selects the gauge-invariant subspace of the Hilbert space, defining the allowed configurations, which consist of pairs of charges connected by electric fluxes (strings) and closed loops of electric flux.

We now map dimer configurations onto these LGT configurations by introducing U$(1)$ gauge fields on every link $l$ and matter fields on each site $x$ of the dual hexagonal lattice, as illustrated in Figs.~\ref{fig:Fig1_Model}(a)-\ref{fig:Fig1_Model}(d). Specifically, we define the electric-field operator as
\begin{equation}
\hat{S}^z_l = (-1)^{s_l} \left[\hat{D}_l -\frac{1}{2} \mathbb{I}\right],
\end{equation}
where $s_l=0$($1$) for links $l$ connecting sites that belong to the same (different) unit cell, as illustrated in Fig.~\ref{fig:Fig1_Model}(b). By construction,  $\hat{S}^z_l$ is a spin-$1/2$ operator, providing a way to assign electric fields to links in a truncated U$(1)$ LGT (quantum link model)~\cite{Chandrasekharan_1997,Wiese_2013}. With this definition, the star-order dimer configuration associated with the $2/3$ (or equivalently $1/3$) atomic phase  corresponds to the state with $\hat{S}^z_l=-1/2$ on all links, which we refer to as the bare vacuum of the LGT. Gauge-field strings are defined as deviations from this reference configuration~\cite{Zeng_1996}: they correspond to links where $\hat{S}^z_l=+1/2$, which in the dimer/atomic picture appear as one-dimensional defects with respect to the ordered pattern.

Matter fields (charges) reside on the sites of the dual lattice, located at the centers of triangles formed by neighboring spins [Fig.~\ref{fig:Fig1_Model}(b)]. In the dimer picture, they correspond to local violations of the dimer constraint, known as monomers~\cite{Rokhsar_1988, Olav_2005, Poilblanc_2006, Ralko_2007, Poilblanc_2010, Schwandt_2010}.
Equivalently, in the atomic or spin representation, these charges appear as frustrated triangles. In the bare vacuum, each triangle is only minimally frustrated, with two spins aligned and one anti-aligned. A configuration in which all three spins align violates this local rule, producing a fully frustrated triangle that maps to a charge excitation in the gauge description. Accordingly, we define the charge operator on a dual-lattice site $x$ as
\begin{equation}
\label{eq:dyn_charge}
     \hat{Q}_{x}/2 = (-1)^{s_x+1} \left[\ket{rrr}_{{\bf i}{\bf j} {\bf k}} \bra{rrr} + \ket{ggg}_{{\bf i}{\bf j} {\bf k}} \bra{ggg}\right],
\end{equation}
where the indices ${\bf i}, {\bf j}$ and ${\bf k}$ label the atomic sites at the vertices of the triangle associated with $x$, and $s_x = 0$ or $1$ indicates which of the two sublattices in the unit cell the site $x$ belongs to [Fig.~\ref{fig:Fig1_Model}(b)].

With these definitions of gauge and matter fields, every dimer configuration corresponds to a U($1$) LGT state $\ket{\psi}$ satisfying a local Gauss-law constraint on the hexagonal lattice 
\begin{equation}
    \hat{G}_x\ket{\psi}=\left[q_x + \frac{(-1)^{s_x}}{2}\right]\ket{\psi},
\end{equation}
where $q_x$ denotes the static charge at site $x$, defined relative to the staggered background charge $(-1)^{s_x}/2$, and the generator of the local U$(1)$ symmetry are
\begin{equation}
   \hat{G}_x =  \nabla_x \hat{S}^z - \hat{Q}_x.
\end{equation} 
Here, $\hat{Q}_x$ denote the dynamical charges, and the lattice divergence of the electric field is defined such that the electric flux flows from positive to negative charges as
\begin{equation}
    \nabla_x \hat{S}^z = (-1)^{s_x+1}\sum_{l\in x} (-1)^{s_l} \hat{S}^z_l.
\end{equation}
Static charges label the different gauge symmetry sectors of the Hilbert space, with the bare vacuum belonging to the sector with $q_x=0, \, \forall x$. Since Gauss's law originates from the construction of the Hilbert space, they are local conserved quantities that cannot be violated dynamically.   

We note that, while every atomic configuration maps onto a LGT configuration that satisfies Gauss’s law, the converse does not hold. In particular, Eq.~\eqref{eq:dyn_charge} shows that dynamical charges can only take even integer values, implying that an even number of strings must be attached to each charge. In the following, we therefore focus on configurations consisting of pairs of strings connecting charges $Q=\pm 2$~[Fig.~\ref{fig:Fig1_Model}(a)]. This situation contrasts with that discussed in Ref.~\cite{Gonzalez_2025}, where a U($1$) LGT emerges on a hexagonal lattice through a different mapping based on Rydberg atoms arranged in a Kagome geometry. In that case, the allowed configurations include single strings connecting charges $Q=\pm 1$. 

Moreover, in our system the set of possible string geometries is also constrained. Nevertheless, unlike in the Kagome mapping, the remaining allowed configurations are sufficient to support transverse fluctuations of the string width~[Fig.~\ref{fig:Fig1_Model}(d)], thereby giving rise to a roughening phase. As we now show in detail, a further important distinction between the two mappings is that, in the present case, plaquette interactions emerge already at first order in the Rydberg Hamiltonian of Eq.~\eqref{eq:hamiltonain_rydberg}, rather than as sixth-order processes.

\subsubsection{U(1) LGT and Hamiltonian processes} 
The processes generated by the Rydberg Hamiltonian in Eq.~\eqref{eq:hamiltonain_rydberg} can induce both the breaking and fluctuation of strings connecting static and dynamical charges, making the system an ideal platform for exploring the interplay between them. We discuss now how the relevant terms of a U$(1)$ quantum link model emerge from this Hamiltonian in different parameter regimes.

In the bare vacuum, first-order processes in the Rabi frequency $\Omega$ create either a `string loop' or a `charged loop', as illustrated in Fig.~\ref{fig:Fig1_Model}(c). The former becomes resonant inside the $2/3$ phase at $\Delta/U \approx 3.27$. 
In contrast, the latter becomes resonant only at a much larger detuning $\Delta/U \approx 6$, at the boundary with the disordered phase. 
Thus, `deep' in the $2/3$ phase, i.e., for $ 3.27\lesssim \Delta/U \lesssim 6$, the bare vacuum coincides with the ground state of the Rydberg model, and the low-energy excitations consist of string loops. As one approaches the DQCP  at $\Delta/U\approx3.158$, these loops proliferate in the vacuum, and the ground state becomes a superposition of many interacting loops. Consequently, while deep in the $2/3$ phase, the vacuum physics can be described using first-order perturbation theory, at lower $\Delta/U$, perturbative calculations fail.
In the following, we focus on the former regime, while the latter will be analyzed numerically in the following sections. 

Deep in the $2/3$ phase, due to Gauss's law, pairs of charges with $Q=\pm 2$ (static or dynamical) are connected by two rigid gauge-field strings, as shown in Fig.~\ref{fig:Fig1_Model}(a). 
In the limit $\Omega=0$, the competition between detuning $\Delta/U$ and interaction strength $U$ gives rise to a linear confining potential: the energy $E_{{\rm ch}}(R)$ of two static charges at a distance $R$, relative to the bare-vacuum energy $E^{(0)}_{{\rm vac }}$, increases linearly with $R$. The resulting confining potential $V_{\Omega=0}(R)=E_{{\rm ch}}(R)-E^{(0)}_{{\rm vac }}$ is given by 
\begin{equation}
    V_{\Omega=0}(R) = \sigma R + 2m ,
    \label{eq:confinig_potential_cl}
\end{equation}
where $\sigma$ is the string tension, and $2m$ is the mass of a pair of static charges. Including interactions up to third-nearest-neighbor, these parameters are given by
\begin{align}
     &\sqrt{3} a \, \sigma \approx \Delta - \left(3 U +9 U_2 -2U_3\right), \\
     & 2m \approx 5\Delta/3 - \left(4 U +10 U_2 +20U_3/3\right),
    \label{eq:sigma_offset_Omega0}
\end{align}
where $U_2=U/27$ and $U_3=U/2^6$ are the second- and third-nearest-neighbor interaction strengths. 

Because the potential increases linearly as the charges are pulled apart, the string connecting them becomes energetically unstable once their separation exceeds $d_* = 2m/\sigma$. For $d \gtrsim d_*$, it is favorable to break the long string and create additional dynamical charge pairs at shorter distances. The value of $d_*$ is controlled by the ratio $\Delta/U$: for large $\Delta/U$, $d_* \to 2$, so long strings always break, whereas as $\Delta/U \to 3.3$, $d_* \to \infty$, and the strings no longer break. This signals a point within the $2/3$ phase where strings connecting static charges can ``freely" fluctuate. This picture, however, changes in the presence of quantum fluctuations. 

In second-order perturbation theory in $\Omega$, the string tension $\sigma$ no longer vanishes within the phase but instead diverges at $\Delta/U \approx 3.27$, the resonant point associated with vacuum string-loop excitations. As illustrated in Fig.~\ref{fig:Fig1_Model}(d), loops can also form inside the string, causing it to break in first order in $\Omega$, or attach to the string, increasing its transverse width in third order in $\Omega$. Together, these processes allow the string to both break and fluctuate. In the following sections, we numerically demonstrate that charges remain confined in the whole $2/3$ phase and that, while strings are rigid deep within the phase, quantum fluctuations near the DQCP drive string roughening [Fig.~\ref{fig:Fig1_Model}(e)].

\subsection{Parameters and geometries} 

In this work, we consider static charges $q=\pm2$ placed at increasing distances $R/a=\sqrt{3}(d+1/3)$, where $d\in \mathbb{N}$ [Fig.~\ref{fig:Fig1_Model}(a)]. These charges are imposed by removing the atoms associated with the corresponding frustrated triangles, a procedure that can be straightforwardly implemented in programmable tweezer arrays~\cite{Gonzalez_2025}. 

In Sec.~\ref{sec:roughening}, we first investigate the equilibrium properties of both the vacuum and static-charge sectors for distances $d=3-9$, scanning the ratio $\Delta/U$ for fixed values of the Rabi frequency $\Omega/U=0.12,0.18$ and $0.21$. Starting `deep' in the $2/3$ phase ($\Delta/U=3.32$), we decrease $\Delta/U$ toward the DQCP, down to $\Delta/U=3.18$. For each parameter set, we compute the ground state of the system using the density matrix renormalization group (DMRG) method. We then compare configurations with and without static charges to extract the width of the string and the confining potential.
In Sec.~\ref{sec:dynamics}, we investigate the dynamics of an initially prepared rigid string separating two static charges at distance $d$, quenching to $\Omega/U = 0.18$ for different values of $\Delta/U$, using the time-dependent variational principle (TDVP) method. In App.~\ref{sec:App_numerics}, we provide further details on the numerics.

In the following, we fix the lattice length along the transverse direction and place the static charges at the transverse boundaries, as shown in Fig.~\ref{fig:Fig2_Width_profile}(a). This effectively freezes the degrees of freedom of the lower flux tube, while the remaining string is free to fluctuate. We do not decrease $\Delta/U$ below $3.18$, as the string then begins to touch the transverse boundaries, introducing strong finite-size effects.  Furthermore, we truncate the interactions to the third-nearest neighbors. In App.~\ref{sec:App_5NN}, we show that extending the truncation to the fifth-nearest neighbors leads only to small quantitative modifications and does not affect our overall results.

\section{String roughening}\label{sec:roughening}
In this section, we study the equilibrium properties of strings that connect static charges within the $2/3$ phase, showing that deep in the phase the strings remain rigid, whereas upon approaching the DQCP, they fluctuate and become rough [Fig.~\ref{fig:Fig1_Model}(e)]. We confirm the roughening quantitatively by first showing that the string width increases with the separation of static charges consistent with a logarithmic scaling, and second that the confining potential acquires a correction to the linear form consistent with the L\"uscher term.

\subsection{Electric flux tube width}
\begin{figure}
    \centering
    \includegraphics[width=\columnwidth]{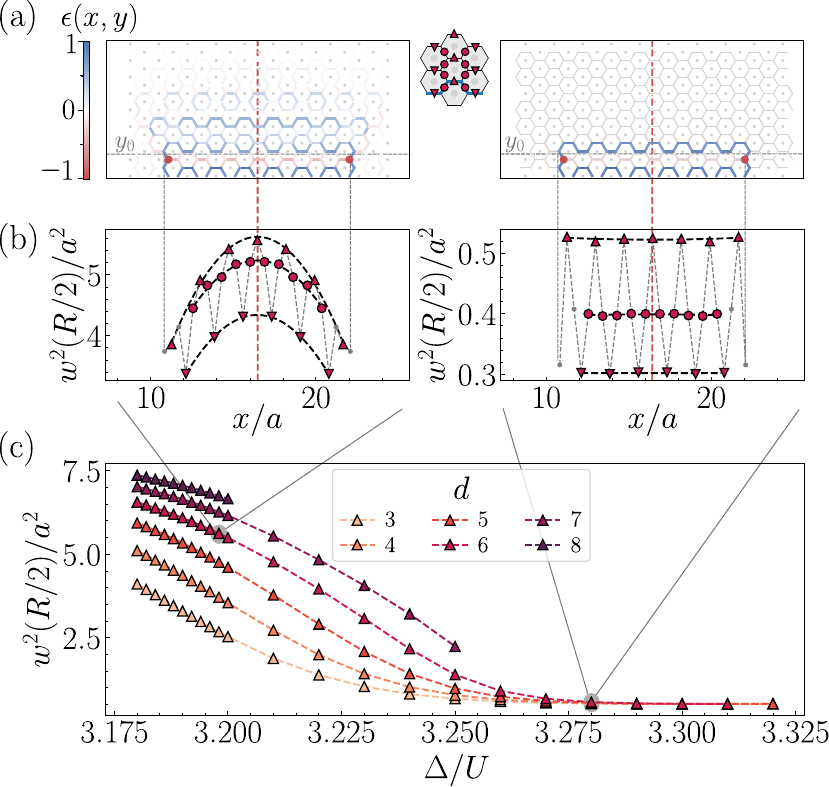}
    \caption{(a)-(b) Illustration of the definition of the string width. The width $w^2(x)$ of the string connecting two static charges separated by a distance $R$ along the $x$-direction, shown in (b), is computed from the string profile in (a) using Eq.~\eqref{eq:w2_x}. Due to the choice of $y_0$, $w^2(x)$ exhibits oscillations, revealing three sublattices. (Left) For small $\Delta/U$, when the string is rough, $w^2(x)$ displays a parabolic shape with maximum at $R/2$. (Right) For large $\Delta/U$, when the string is rigid, $w^2(x)$ remains constant within each sublattice. (c) String width at $R/2$ as a function of $\Delta/U$ for one sublattice of (b), estimated from a parabolic fit in $x$ (corresponding black dashed line in (b)). For large $\Delta/U$, $w^2(R/2)$ remains constant as $R$ increases, while, for small $\Delta/U$, $w^2(R/2)$ increases with $R$.  Results obtained using a finite MPS with bond dimension $\chi=900$. }
    \label{fig:Fig2_Width_profile}
\end{figure}

While the transverse width of a rigid string remains constant as the separation $R$ between static charges increases, it diverges logarithmically for a rough string due to transverse fluctuations.
In particular, the string width at half the distance between the charges is expected to grow as
\begin{equation}
    w^2(R/2) = \frac{v}{2\pi\sigma} \log{\left(\frac{R}{R_0}\right)}, \:\:\ \text{for} \:\:\: R\gg R_0,
    \label{eq:width_log_scaling}
\end{equation} 
where $R_0$ is a characteristic length scale, and $v$ is a (non-universal) velocity factor that accounts for the mapping from the standard Lorentz-invariant expression~\cite{Luscher_1981_1, Caselle_1996, Gliozzi_2010} to our Hamiltonian framework.

We verify the expected logarithmic growth considering static charges separated by increasing distances $R$, for values of $\Delta/U$ approaching the DQCP from deep in the $2/3$ phase. Following Refs.~\cite{Luscher_1981_1, DiMarcantonio_2025}, we define the width at half distance between the charges, $w^2=w^2(x=R/2)$, as illustrated in Figs.~\ref{fig:Fig2_Width_profile}(a)-\ref{fig:Fig2_Width_profile}(b). We consider 
\begin{equation}
    \epsilon_l = \langle\hat{S}^z_l\rangle_{\rm ch}  - \langle\hat{S}^z_l\rangle_{\rm vac},
    \label{eq:energy_density}
\end{equation}
which quantifies the variation in the electric field due to the presence of strings. Here, the subscripts `ch' and `vac' refer to the configuration with static charges and the vacuum, respectively. The flux-tube width along a fixed cut at  $x$ is then obtained from
\begin{equation}
    w^2(x) = \frac{\sum_{y>y_0} \, (y-y_0)^2 \epsilon(x,y) }{ \sum_{y>y_0} \, \epsilon(x,y)},
    \label{eq:w2_x}
\end{equation}
where $(x,y)$ is the coordinate associated with the center of the link $l$ in the hexagonal lattice, and $y_0$ is a reference position (chosen to be $y_0=y_{\rm ch} + 0.25a$ to exclude links within the rigid tube). Based on this definition, due to the hexagonal structure of the lattice, we distinguish three different sublattices. To extract the width at half distance between the charges $w^2(R/2)$, we fit a parabolic function to the profile $w^2(x)$ for each sublattice, as illustrated in Fig.~\ref{fig:Fig2_Width_profile}(b), and take the value of the  fitted function at $x=R/2$ as the estimate.

\begin{figure}
    \centering
    \includegraphics[width=\columnwidth]{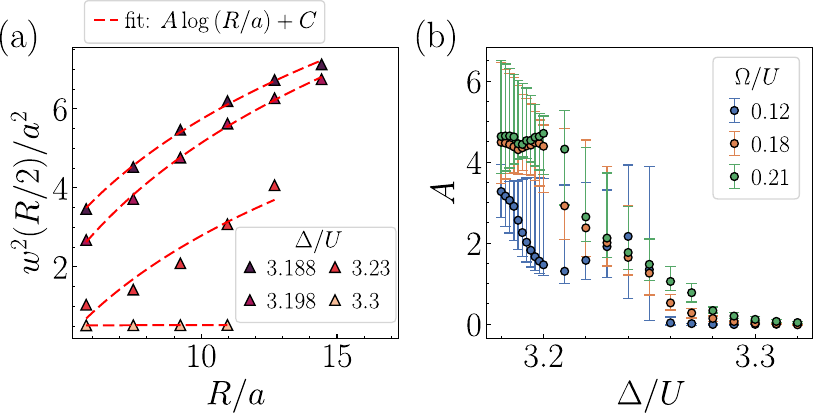}
    \caption{(a) String width $w^2(R/2)$ as a function of $R$ for three representative $\Delta/U$ in Fig.~\ref{fig:Fig2_Width_profile}(c). For small $\Delta/U$, e.g., $\Delta/U=3.188$ and $3.198$, $w^2(R/2)$ increases with R, consistent with the expected logarithmic dependence; for large $\Delta/U$, e.g., $\Delta=3.3$, it remains constant; while for intermediate values there is a crossover between the two behaviors. Red lines indicate fits to $A\log{R}+C$. (b) Estimate of the fit parameter $A$ as a function of $\Delta/U$ for different $\Omega/U$, extracted via the analysis described in App.~\ref{sec:App_procedure_fit}.}
    \label{fig:Fig3_width_log}
\end{figure}

As shown in Fig.~\ref{fig:Fig2_Width_profile}(c) for one sublattice, deep in the $2/3$ phase (gray region) the string width remains unchanged with increasing $R$. For smaller values of $\Delta/U$, however, the width begins to increase. In particular, as illustrated in  Fig.~\ref{fig:Fig3_width_log}(a), after a crossover region displaying different behaviors at short and long distances, the growth becomes increasingly consistent with the logarithmic form $A\log{(R)}+C$ ($A\neq0$) as the system approaches the DQCP. 

This trend is quantified by the fit parameter 
$A$, shown as a function of $\Delta/U$ in Fig.~\ref{fig:Fig3_width_log}(b). While $A$ is compatible with zero deep in the $2/3$ phase (gray region), it becomes finite as $\Delta/U$ decreases. To obtain a final estimate of $A$ and its associated error, we performed a systematic error analysis considering fits over all possible intervals $[R_{\rm min}, R_{\rm max}]$ and sublattices, since variations in these choices constitute the primary sources of uncertainty (see details in App.~\ref{sec:App_procedure_fit})~\cite{Banuls_2013}. We note that the larger uncertainties at small $\Delta/U$ arise from including sublattices other than those shown in Fig.~\ref{fig:Fig3_width_log}(a), for which the logarithmic behavior is less clear (see App.~\ref{sec:App_w2}). We next investigate the confining potential between static charges to establish a connection between the coefficient $A$ and the string tension $\sigma$ [Eq.~\eqref{eq:width_log_scaling}].

\subsection{Confining potential and universal correction}
The energy associated with a rigid string connecting a pair of charges grows linearly with their separation $R$, i.e., $V(R)\sim \sigma R$, where $\sigma$ is the string tension. For sufficiently large $R$, the formation of a long string becomes energetically unfavorable, leading to its breaking and the creation of additional charge pairs at shorter distances; beyond this point, $V(R)$ no longer increases with $R$. 
In addition, for a rough string, transverse fluctuations introduce a universal correction to the linear potential:
\begin{equation}
    V(R) = \sigma R -   \underbrace{v \, \gamma_0}_{\gamma} \frac{1}{R} + \text{const.},
    \label{eq:V(R)_gamma}
\end{equation}
where the velocity factor $v$ relates the universal L\"usher term $\gamma_0=\pi (D-2)/24$ of the Lorentz-invariant formulation~\cite{Luscher_1981}---determined solely by the space-time dimension $D$ ($D=3$ in our case)--- to the corresponding correction in our Hamiltonian formulation, $\gamma=v \gamma_0$.

\begin{figure}
    \centering
    \includegraphics[width=\columnwidth]{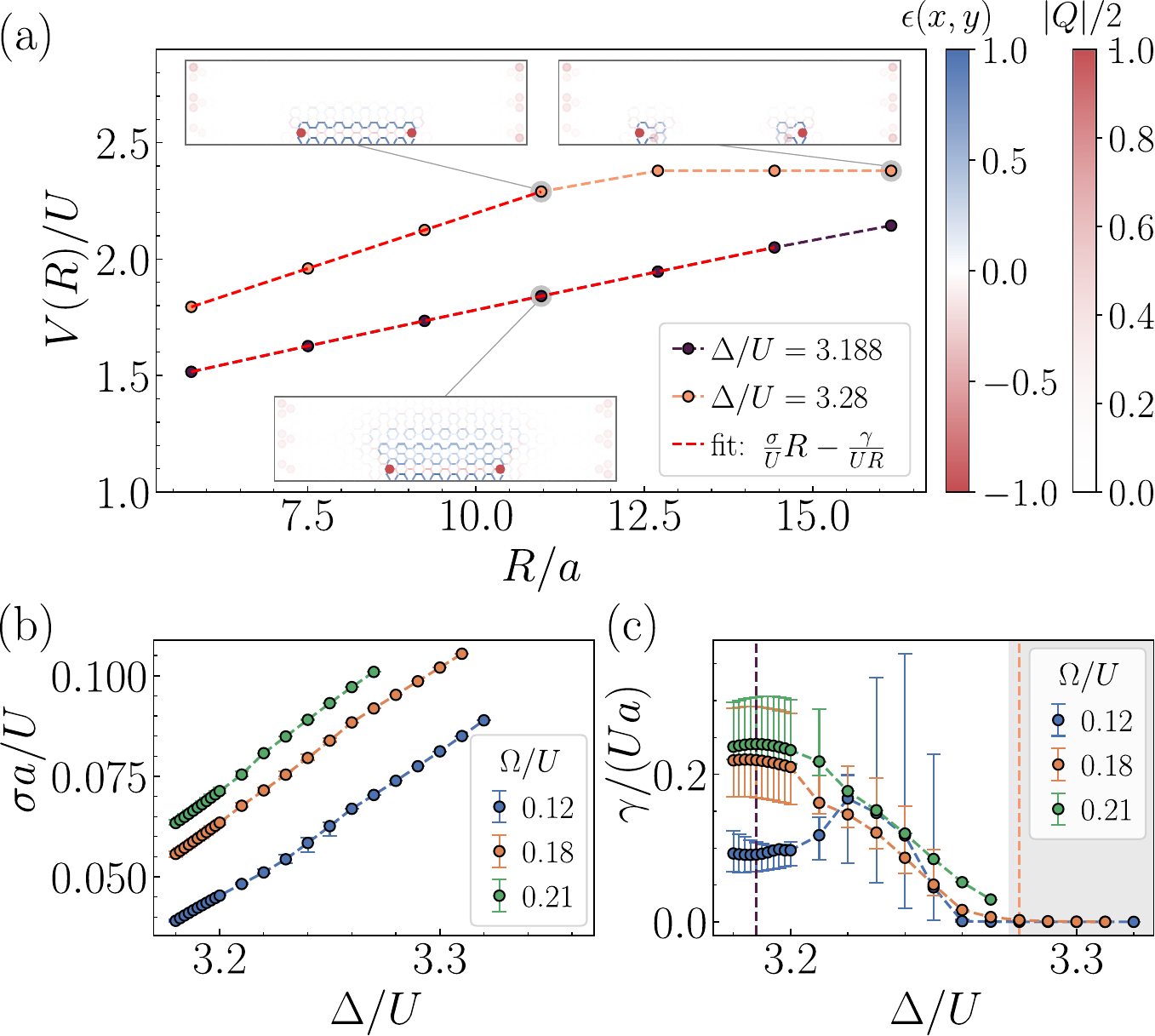}
    \caption{(a) The confining potential $V(R)$ as a function of the distance $R$ between two static charges, shown for two representative values of $\Delta/U$. Red dashed lines indicate fits of the potential according to Eq.~\eqref{eq:V(R)_gamma}. Insets: examples of corresponding string profiles $\epsilon(x,y)$  [Eq.~\eqref{eq:energy_density}] along with the static and dynamical charges $|Q|$. (b)-(c) Fit parameters of $V(R)$ as function of $\Delta/U$ for three values of $\Omega/U$, extracted via the analysis described in App.~\ref{sec:App_procedure_fit}. (b) The string tension $\sigma$ decreases with decreasing $\Delta/U$, i.e., when approaching the DQCP. (c) The $\gamma$ correction vanishes for large $\Delta/U$ (gray region) and becomes finite near the DQCP. The dashed vertical lines indicate the values of $\Delta/U$ used in (a). Results obtained using a finite MPS with bond dimension $\chi=900$.}
    \label{fig:Fig4_potential}
\end{figure}

As illustrated in Fig.~\ref{fig:Fig4_potential}(a), the potential $V(R)=E_{\rm ch}(R)-E_{\rm vac}$ increases with $R$, with a slope (string tension) that qualitatively decreases for lower $\Delta/U$, causing the string to break at shorter distances. To quantitatively extract the string tension $\sigma$ and the correction $\gamma=v \gamma_0$, we fit the potential according to Eq.~\eqref{eq:V(R)_gamma} (see App.~\ref{sec:App_gamma} for a comparison with a purely linear fit).
In Figs.~\ref{fig:Fig4_potential}(b)-\ref{fig:Fig4_potential}(c), we show  $\sigma$ and $\gamma$ as functions of $\Delta/U$.

Deep in the $2/3$ phase, the results agree with the classical $\Omega=0$ limit discussed in Sec.~\ref{sec:model}: $\sigma$ increases linearly with $\Delta/U$, and there is no correction to the linear potential ($\gamma \approx 0$), as indicated by the gray area in panel (c). For $\Delta/U \lesssim 3.28$, the results deviate from these classical expectations. While $\sigma$ is expected to vanish at $\Delta/U \approx 3.3$ for $\Omega=0$, quantum fluctuations stabilize the string, resulting in a finite $\sigma$ across the entire $2/3$ phase, accompanied by a nonzero correction $\gamma \neq 0$ upon approaching the DQCP. (As for the string width, the fit parameters and their uncertainties were estimated via the systematic analysis detailed in App.~\ref{sec:App_procedure_fit}.) To confirm that the logarithmic increase of the string width and the finite $\gamma$ correction emerging upon approaching the DQCP are associated with the roughening phenomena, we next extract the universal value $\gamma_0$, which is predicted to be $\pi/24$ for a rough string in $D=3$ dimensions.

To relate the fit parameter $\gamma$ to the universal value $\gamma_0$, we eliminate the unknown velocity conversion factor $v$ in Eq.~\eqref{eq:V(R)_gamma} considering the relation between the parameter $A$ that characterizes the logarithmic growth of the string width and the string tension $\sigma$, $A=v/(2\pi\sigma)$ [Eq.~\eqref{eq:width_log_scaling}]. We obtain 
\begin{equation}
    \gamma_0 = \frac{1}{2\pi} \frac{\gamma}{\sigma} \frac{1}{A}.
    \label{eq:gamma_0}
\end{equation}

\begin{figure}
    \centering
    \includegraphics[width=\columnwidth]{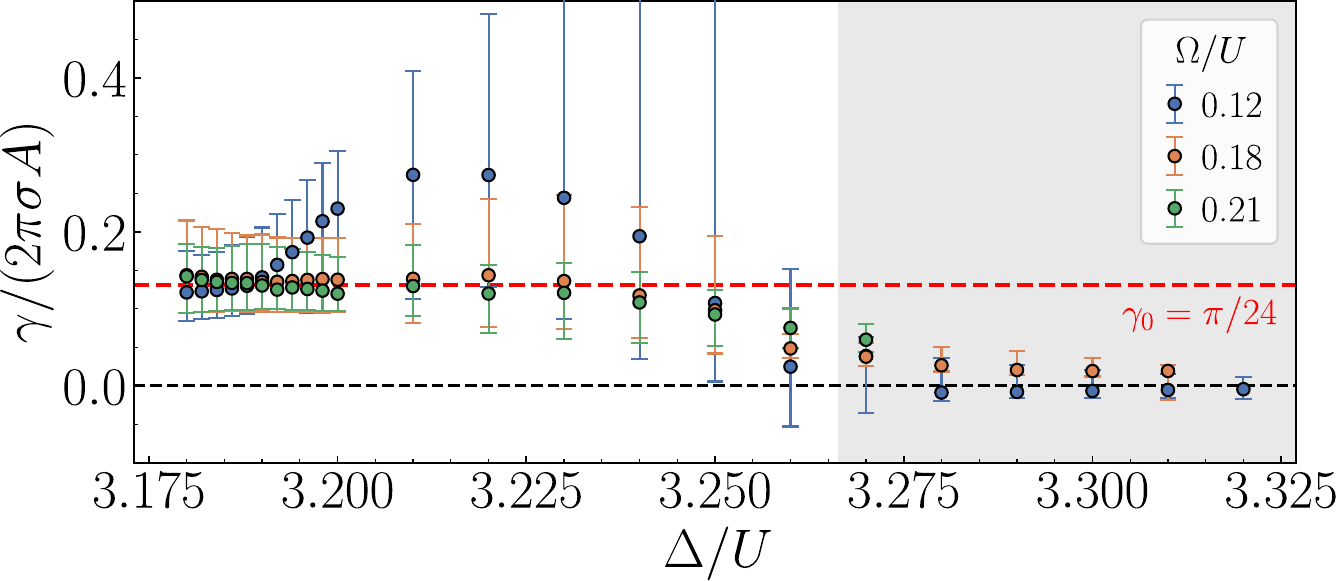}
    \caption{Universal L\"uscher correction to the linear confining potential as a function of $\Delta/U$ for three values of $\Omega/U$, obtained by combining the fit parameters of the string width $A$ [Fig.~\ref{fig:Fig3_width_log}(b)] and confining potential, $\sigma$ and $\gamma$, [Fig.~\ref{fig:Fig4_potential}(b) and (c)] according to Eq.~\eqref{eq:gamma_0}, as detailed in App.~\ref{sec:App_gamma}. The dashed horizontal line indicate the  universal value $\gamma_0=\pi/24$ for a rough string in ($2+1$)D.}
    \label{fig:Fig5_gamma0}
\end{figure}

To estimate $\gamma_0$, we therefore combine the fit parameters of the confining potential and the growth of the string, which are a priori independent, in this manner. As shown in Fig.~\ref{fig:Fig5_gamma0}, the resulting quantity becomes compatible with the universal value $\gamma_0=\pi/24$ upon approaching the DQCP, independently of $\Omega/U$ (see App.~\ref{sec:App_procedure_fit} and App.~\ref{sec:App_gamma} for details on its estimation). This, together with the logarithmic divergence of the string width, signals the onset of string roughening. 

While string roughening is well established in a variety of lattice gauge theories~\cite{Hasenfratz_1981, Munster_1981, Munster_1981_1, Drouffe_1981, Drouffe_1981_1}, it was a priori unclear whether our model based on Rydberg atoms would exhibit this universal behavior. Our results  demonstrate that string roughening can appear in an experimentally accessible system, paving the way for exploring roughening phenomena in analog quantum simulators.

\section{Quench dynamics}\label{sec:dynamics}
Aside from exploring equilibrium properties for different values of $\Delta/U$ and $\Omega/U$, which can be done via adiabatic state preparation, current experiments could also investigate quench dynamics, probing the interplay between string breaking and string roughening. In a quench protocol, the system is initialized (either via local detuning or adiabatic preparation) in the ground state of an initial Hamiltonian and then driven out of equilibrium by a sudden change of the Hamiltonian parameters.

\begin{figure}
    \centering
    \includegraphics[width=\columnwidth]{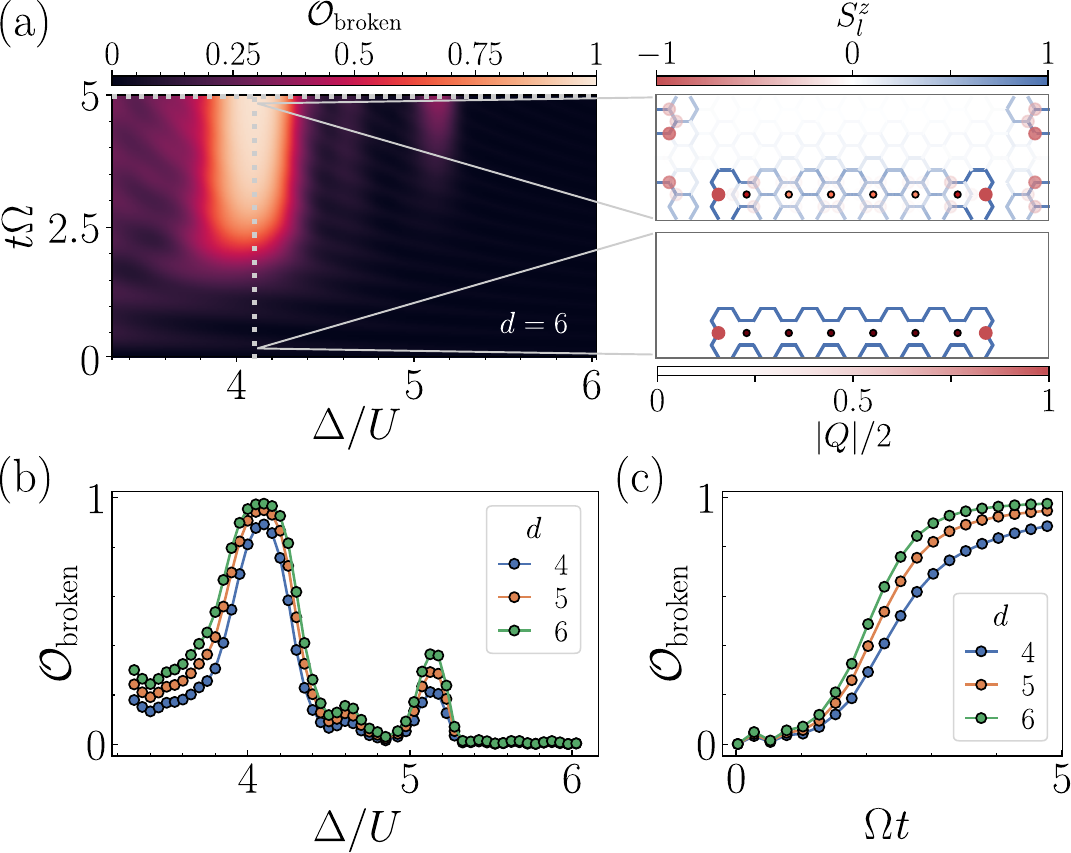}
    \caption{(a) String breaking observable $\mathcal{O}_{\rm broken}$ [Eq.~\eqref{eq:O_broken}] as a function of $\Delta/U$ and time $\Omega t$. Insets show the electric field $\langle \hat{S}^z_l\rangle$ and the corresponding dynamical and statical charges, at the  initial time $\Omega t=0$ and the final time $\Omega t=5$ for $\Delta/U\approx4.1$. (b)  $\mathcal{O}_{\rm broken}$ at the final time displays a resonant peak at $\Delta/U\approx4.1$, which broadens with increasing separation between the static charges $d$. (c) $\mathcal{O}_{\rm broken}$ at $\Delta/U\approx4.1$ grows in time and plateaus at its maxim value after $t\Omega\approx3.6$. Results obtained using a finite MPS with bond dimension $\chi=100$.}
    \label{fig:Fig6_dynamics_breaking}
\end{figure}

We first investigate string breaking starting from the classical rigid string configuration, quenching to $\Omega/U = 0.18$ for values of $\Delta/U$ far from the DQCP. In this regime, transverse fluctuations are strongly suppressed, allowing us to truncate the lattice in the transverse direction to reduce computational cost. To detect whether the rigid string breaks, we monitor the time evolution of 
\begin{equation}
    \mathcal{O}_{\rm broken} = 1- \prod_{i\in S_{\rm up}} \langle \hat{n}_{i} \rangle,
    \label{eq:O_broken}
\end{equation}
where the product runs over atoms in the rigid-string region that are initially in state $\ket{r}$ [Fig.~\ref{fig:Fig6_dynamics_breaking}(a)]. By construction, $\mathcal{O}_{\rm broken}=0$ for a rigid string and $\mathcal{O}_{\rm broken}=1$ for broken strings, irrespective of how the breaking occurs.

We expect to observe dynamical string breaking when we quench to values of $\Delta / U$ for which the initial unbroken string is energetically resonant with broken string configurations~\cite{Gonzalez_2025}. As shown in Fig.~\ref{fig:Fig6_dynamics_breaking}(a) for two static charges at distance $d=6$,  $\mathcal{O}_{\rm broken}$ increases substantially in time only for $\Delta/U$ in the range $3.8-4.4$. Outside this interval, $\mathcal{O}_{\rm broken}$ remains low, indicating that the rigid string remains intact. As shown in Fig.~\ref{fig:Fig6_dynamics_breaking}(b) at the final time, the resonance broadens with increasing charge separation $d$, reflecting the growing number of classical broken-string configurations that become resonant.
Finally, in Fig.~\ref{fig:Fig6_dynamics_breaking}(c), we show $\mathcal{O}_{\rm broken}$ for $\Delta/U$ corresponding to the peak at the final time, highlighting its initial growth and subsequent saturation.  
 
\begin{figure}
    \centering
    \includegraphics[width=\columnwidth]{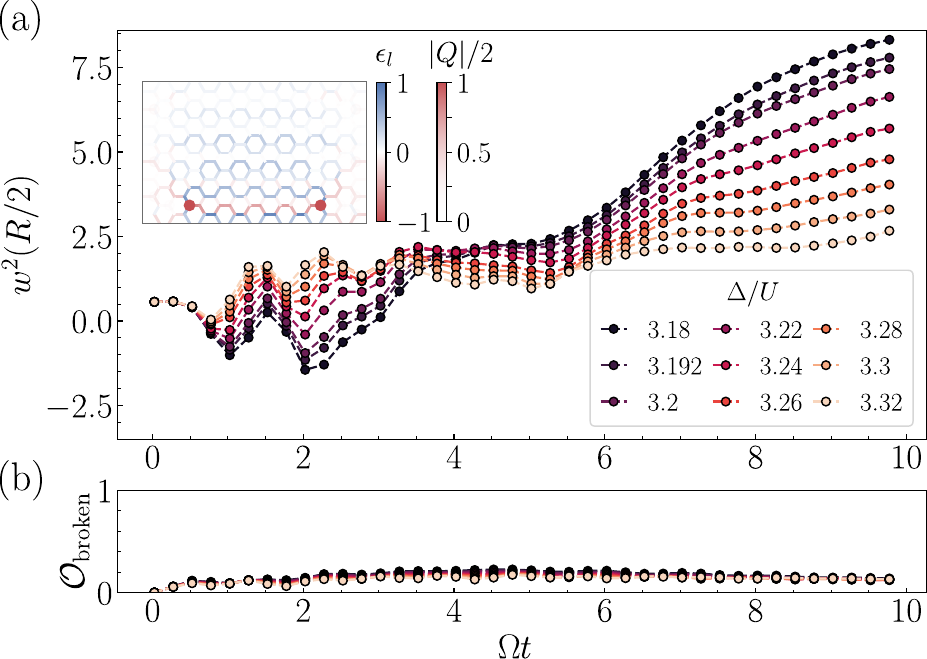}
    \caption{ (a) String width $w^2(R/2)$ at half distance between two static charges at distance $d=4$, as a function of time $\Omega t$ for different values of $\Delta/U$. The inset displays, for the smallest $\Delta/U$ at final time, the string profile $\epsilon_l$ [Eq.~\eqref{eq:energy_density}] along with the corresponding static and dynamical charges. (b) String-breaking observable $\mathcal{O}_{\rm broken}$, corresponding to the data in panel (a), as a function of time.  Results obtained using a finite MPS with bond dimension $\chi=300$ (see App.~\ref{sec:App_dynamics} for comparison with data at lower $\chi$).}
    \label{fig:Fig7_dynamics_w2}
\end{figure}

We investigate now dynamical string fluctuations by considering two static charges at distance $d=4$ and quenching the rigid string closer to the DQCP. As shown in Fig.~\ref{fig:Fig7_dynamics_w2}(a), after an initial transient dominated by vacuum fluctuations, the string width $w^2(R/2)$ begins to grow in time. As $\Delta/U$ is tuned closer to the transition, this growth becomes more pronounced, reflecting increasingly strong transverse fluctuations. Furthermore, during these dynamics the string breaks only with low probability, as indicated by the small value of $\mathcal{O}_{\rm broken}$, which remains below 0.2, as illustrated in Fig.~\ref{fig:Fig7_dynamics_w2}(b). 

Our results show that, for short string lengths, width fluctuations and string breaking emerge as distinct dynamical phenomena across different parameter regimes. At equilibrium, however, these two effects occur simultaneously for strings connecting static charges at sufficiently large separations ($d \gtrsim 7$), as illustrated in Fig.~\ref{fig:Fig2_Width_profile}(c). In this regime, the onset of string breaking approaches the roughening crossover as $d$ increases. Although simulating the real-time dynamics of such long, strongly fluctuating strings lies beyond current numerical capabilities, state-of-the-art quantum simulators can access this parameter regime: experiments with hundreds of Rydberg atoms can directly probe the non-equilibrium dynamics, opening the door to exploring the interplay between string breaking and roughening in a ($2+1$)D U($1$) LGT.

\section{Conclusions and outlook}\label{sec:conclusions}
In summary, we have shown that Rydberg atoms arranged on a triangular array can be used as a quantum simulator to explore string roughening phenomena, which is relevant in high-energy and condensed matter physics.
In  ($2+1$)D or higher dimensions, strings of gauge fields that confine excitations can fluctuate in the transverse direction, leading to roughening behavior. Such fluctuations are typically driven by plaquette terms, which are challenging to implement in current analog simulators. Here, we demonstrate that string roughening emerges naturally when approaching a DQCP in the phase diagram of the Rydberg Hamiltonian, which separates the $1/3$ and the $2/3$ ordered phases in the system.

By mapping configurations of atoms to configurations of a U($1$) LGT, we show that charge excitations are confined by strings. Deep in the phase, these strings remain rigid, while near the DQCP they become rough. We quantify this behavior by analyzing the growth of both the string width with charge separation and the confining potential. Combining these observations, we extract the universal Lu\"scher correction, finding results compatible with the expected value for a rough string in ($2+1$)D. 

We further investigate quench dynamics, demonstrating that quantum simulators can probe both string roughening and string breaking, as well as their interplay, in both equilibrium and non-equilibrium settings. These results provide concrete guidance for future experimental studies and highlight the potential of neutral-atom platforms to explore confinement phenomena in lattice gauge theories.

\section{Acknowledgments}
L.B. thanks Gabriele Callieri for helpful discussions. This work is supported by the European Union’s Horizon Europe research and innovation program under Grant Agreement No.~101113690 (PASQuanS2.1), the ERC Starting grant QARA (Grant No.~101041435), the Austrian Science Fund (FWF) (Grant No. DOI 10.55776/COE1), and the ERC Starting grant QS-Gauge (Grant No.~101220401).
Views and opinions expressed are however those of the author(s) only and do not necessarily reflect those of the European Union or the European Research Council Executive Agency. Neither the European Union nor the granting authority can be held responsible for them.
D.G.-C. acknowledges financial support through the Ramón y Cajal Program (RYC2023-044201-I), financed by MICIU/AEI/10.13039/501100011033 and by the FSE+. Tensor network calculations were performed using the TeNPy Library~\cite{tenpy2024}. The results have been obtained using the LEO HPC infrastructure of the University of Innsbruck.

\appendix

\section{Numerical methods}\label{sec:App_numerics}

The numerical results presented in this work are obtained using tensor-network methods as implemented in the Python package TeNPy~\cite{tenpy2024}. 
We determine the equilibrium ground state of the Rydberg atom arrays [Eq.~\eqref{eq:hamiltonain_rydberg}], both with and without static charges, using the density-matrix renormalization group (DMRG) algorithm. A one-dimensional matrix product state (MPS) ansatz is employed by contracting the tensors along the traversal-$y$ direction. The calculations use a bond dimension of $\chi=900$ and are converged in energy below error $10^{-9}$. 
For dynamical simulations, we use the time-dependent variational principle (TDVP) algorithm with a time step $\delta t=0.025/\Omega$ and a maximum bond dimension of $\chi=300$.
In the main text, the range of Rydberg interactions is truncated at a distance $2a$, including up to the third-nearest-neighbor interactions. In App.~\ref{sec:App_5NN}, we extend the interaction range up to the fifth-nearest neighbors.

\section{Procedure for the estimation of the fit parameters}\label{sec:App_procedure_fit}
Here, we provide details of the procedure used to estimate the fit parameters, and their uncertainties, for the logarithmic increase of the string width $w^2(R/2)$ as a function of distance [Fig.~\ref{fig:Fig3_width_log}(b)], as well as for the dependence of the confining potential $V(R)$ [Figs.~\ref{fig:Fig4_potential}(b)-\ref{fig:Fig4_potential}(c)].

As discussed in the main text, the main source of systematic uncertainty arises from the choice of the fitting range $[R_{\rm min}, R_{\rm max}]$. In particular, the theoretical predictions in Eq.~\eqref{eq:width_log_scaling} and Eq.~\eqref{eq:V(R)_gamma} are expected to hold only at sufficiently large distances, which motivates excluding short-distance data from the fits. At the same time, large distances can be more affected by boundary effects, especially when approaching the DQCP due to transverse fluctuations of the string. Since there is no fully systematic way to select this interval, we instead consider all possible fitting ranges that include at least four data points. The resulting statistical distribution over these different fitting ranges provides a meaningful estimate of the fit parameters and their uncertainties.

In particular, we consider different sets of consecutive points 
$\{(R_{i_\alpha}, q_{i_\alpha})\}_{i_\alpha=1}^{n_\alpha}$ where $q$ denotes the observable under analysis (e.g., the confining potential or the string width), and  $i_{\alpha}$ labels the element in the $\alpha$-th set. Each set contains $n_{\alpha}\geq n_{\mathrm{min}} = 4$ points.
Each subset $\alpha$ is independently fitted to the target function $g(R; \{p\}_j)$,
where $\{p\}_j$ denotes the corresponding set of fitting parameters, e.g., $p \in \{\sigma, \gamma, \text{const.}\}$ in the case of the confining potential.
For each parameter $p_j$, this procedure yields an estimate $\mathcal{P}_{j, \alpha}$
associated with each subset $\alpha$. 
The final estimate of each parameter $\bar{p}_j$ is defined as the median of the distribution of $\{\mathcal{P}_{j, \alpha}\}_{\alpha}$. The associated uncertainties are obtained from the quantiles corresponding to $\sigma_{-} = 0.1585$ and $\sigma_{+} = 0.8415$, which correspond to the lower and upper bounds of a central $68\%$ interval (i.e., the $\pm 1\sigma$ range for a Gaussian distribution). The uncertainty of $\bar{p}_j$ is then defined by the interval $\left[\bar{p}_k-\sigma_{-}, \bar{p}_k+\sigma_{+}\right]$.

For the estimation of the parameter $A$ in Fig.~\ref{fig:Fig3_width_log}, we extend this procedure by combining results from different sublattices, whose choice introduces an additional source of systematic error. We consider sets of consecutive points $\{(R^{k}_{i_\alpha}, q^{k}_{i_\alpha})\}_{i_\alpha=1}^{n_\alpha}$, now additionally labeled by the sublattice index $k=0,1,2$. Applying the same fitting procedure to each subset, we obtain a set of estimates of $A$, $\{\mathcal{A}^{k}_{j, \alpha}\}_{\alpha}$, for each sublattice. To obtain a conservative estimate of $A$, we combine all these results by taking the median of the full distribution $\{\mathcal{A}^{k}_{j, \alpha}\}_{\alpha, k}$. The corresponding uncertainties are defined as previously. In App.~\ref{sec:App_w2}, we provide a comparison of the results for each sublattice analyzed separately.

\begin{figure}
    \centering
    \includegraphics[width=\columnwidth]{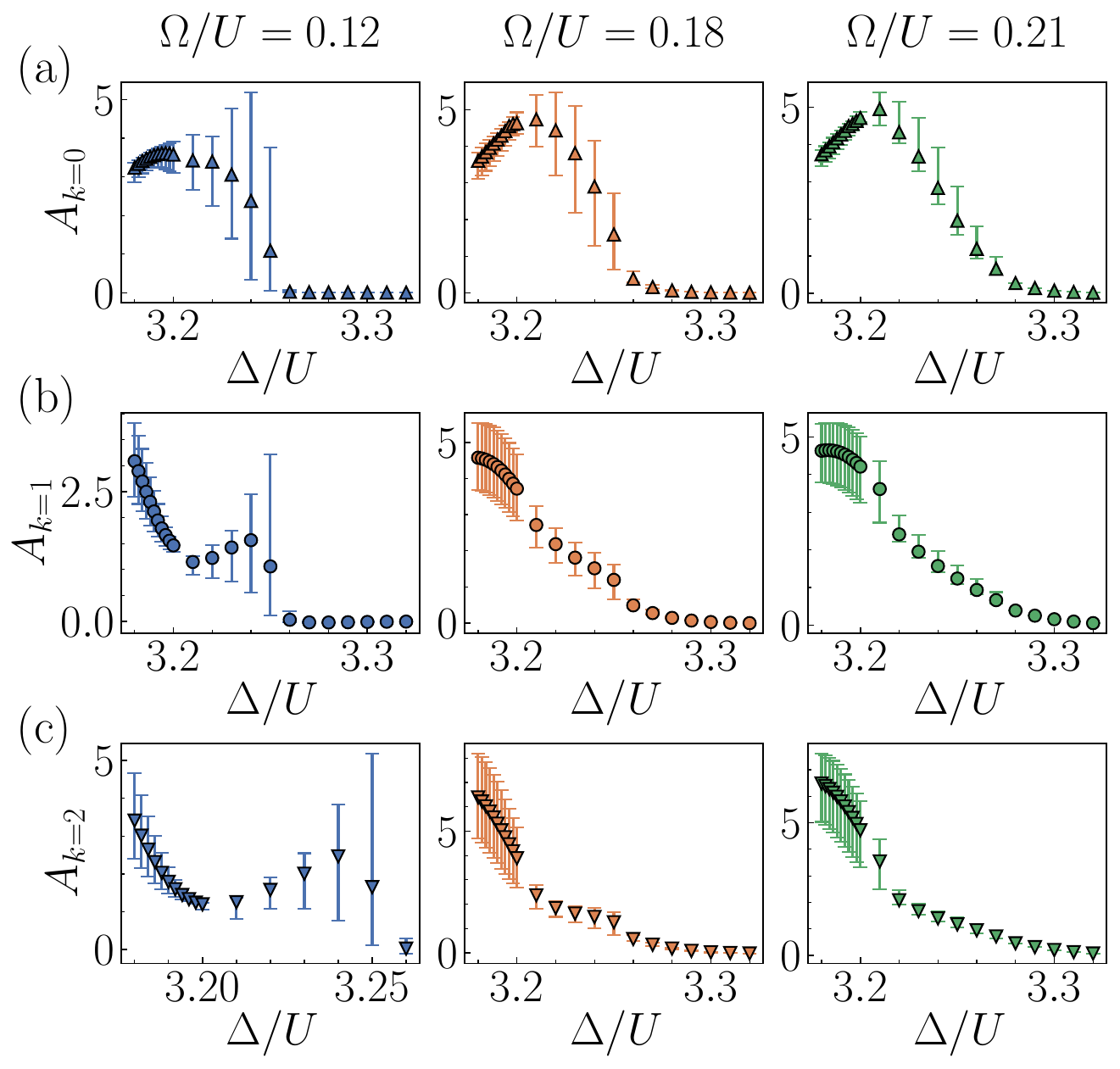}
    \caption{Estimates of the fit parameter $A$, characterizing the logarithmic scaling of the string width, as a function of $\Delta/U$. Rows correspond to different sublattices labeled by $k$, and columns to different values of $\Omega/U$. The parameter $A$ is extracted separately for each sublattice using the procedure described in App.~\ref{sec:App_procedure_fit}.}
    \label{fig:App_A_sublattices}
\end{figure}

\section{Further details on the string width}\label{sec:App_w2}
Here, we provide further details on the analysis of the string width, complementing Fig.~\ref{fig:Fig2_Width_profile}. As explained in the main text, the string width at half distance $w^2 (R/2)$ (for unbroken strings) is obtained by computing the 1D profile $w^2(x)$ using Eq.~\eqref{eq:w2_x}. Due to the hexagonal lattice geometry, the sites along the $x$ axis fall into three distinct sublattices---e.g. the transverse coordinate $(y - y_0)$ takes different values on the three sublattices.  We therefore fit $w^2(x)$ for the sites belonging to each of the three sublattices separately using a parabolic fit and use these fits to evaluate $w^2_k(R/2)$, with $k=0,1,2$ labeling the three sublattices. The indexing is chosen such that $k=0$ corresponds to the sublattice with the largest transverse displacement for a classical rigid string, followed by $k=1$ and $k=2$, as represented in Fig.~\ref{fig:Fig2_Width_profile}(b) by the markers $\triangle$, $\circ$, and $\triangledown$, respectively.

\begin{figure}
    \centering
    \includegraphics[width=\columnwidth]{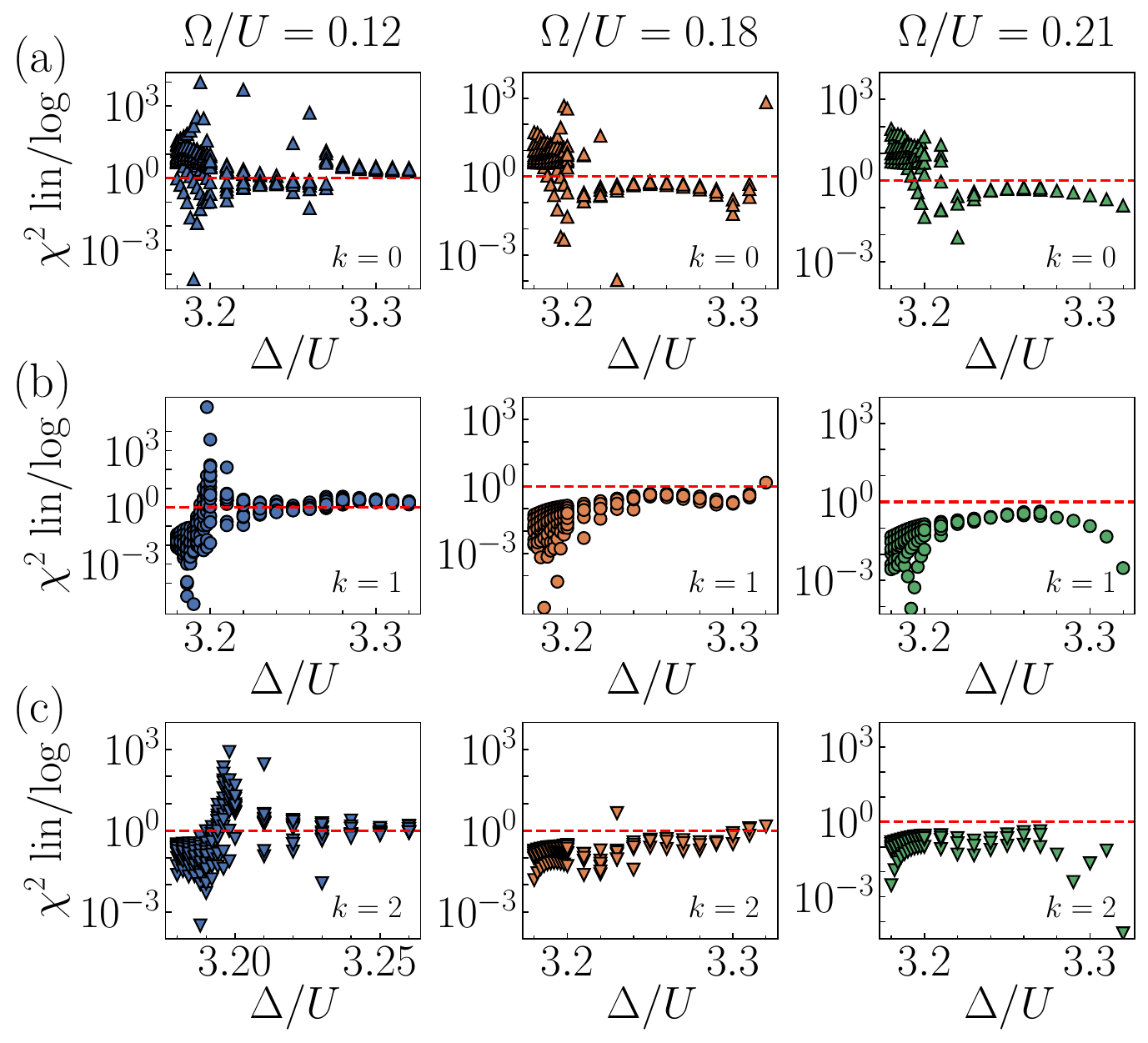}
    \caption{Comparison between linear and logarithmic fits of $w^2(R/2)$, quantified by the ratio of their respective chi-squared values $\chi^2$, evaluated for all choices of the fitting range considered in Fig.~\ref{fig:App_A_sublattices}. Ratios larger than 1 (horizontal dashed line) indicate that the logarithmic fit provides a better description of the data. Rows correspond to different sublattices labeled by $k$, and columns to different values of $\Omega/U$.}
    \label{fig:App_w2_fit_chi2linlog}
\end{figure}

\paragraph*{Estimation of $A$ for logarithmic width growth---}
For a rough string, the width $w^2(R/2)$ increases logarithmically with $R$. We therefore fit the data using the function $A \log R + C$ and estimate the parameters using the systematic procedure described in App.~\ref{sec:App_procedure_fit}.
The final estimate of $A$ and the corresponding uncertainty is illustrated in Fig.~\ref{fig:App_A_sublattices} for the three sublattices considered separately [c.f. Fig.~\ref{fig:Fig3_width_log} in the main text, which combines all sublattices]. For the $k=0$ sublattice and small $\Delta/U$, the uncertainty is smaller, as the logarithmic behavior is more pronounced and better captured by the fit than in the other sublattices. The larger uncertainty in Fig.~\ref{fig:Fig3_width_log} arises from combining the three sublattices.

\paragraph*{Logarithmic growth vs. linear growth---}
We now compare the logarithmic hypothesis with a linear one.
In Fig.~\ref{fig:App_w2_fit_chi2linlog}, we compare the two fitting forms using the ratio of their chi-squared $\chi^2$ values for all fitted intervals $\alpha$. Ratios greater than 1 indicate that the logarithmic fit works better than the linear one. As noted above, the result depends on the sublattice: for the $k=0$ sublattice, the logarithmic form is favorable for all $\Delta/U\lesssim3.2$, whereas for $k=1$ and $2$, it becomes preferable only for a narrow range of $\Delta/U$ for $\Omega/U=0.12$. This sublattice dependence may arise from different boundary effects experienced by the three sublattices as a result of the underlying lattice geometry. In the continuum limit, where the lattice spacing $a$ becomes negligible, these effects are expected to vanish, yielding consistent behavior across all sublattices.

\section{Further details on the confining potential and the universal L\"usher term}\label{sec:App_gamma}
Here, we provide further details on the analysis of the confining potential and on the universal correction $\gamma_0$, complementing Fig.~\ref{fig:Fig4_potential} and Fig.~\ref{fig:Fig5_gamma0} of the main text. 

\begin{figure}
    \centering
    \includegraphics[width=\columnwidth]{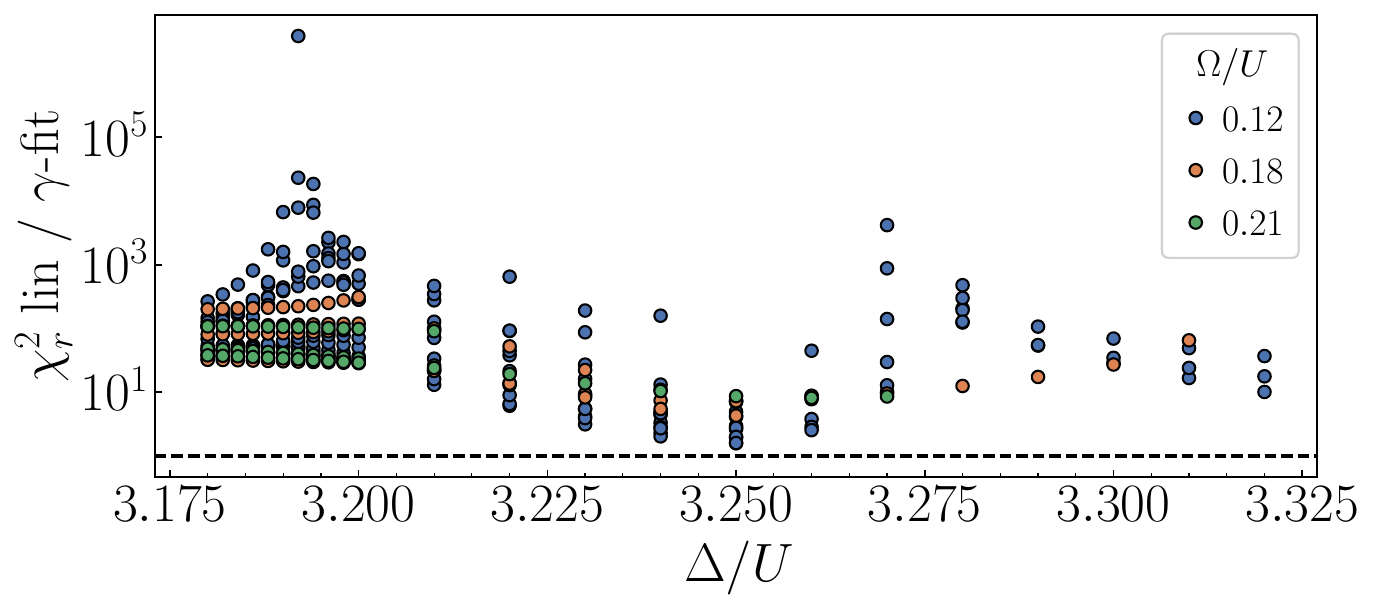}
    \caption{Comparison between linear fits and fits including the $\gamma$ correction [Eq.~\eqref{eq:V(R)_gamma}] of the confining potential $V(R)$, quantified by the ratio of their respective reduced chi-squared values $\chi^2_r$, evaluated for all choices of the fitting range considered in Fig.~\ref{fig:Fig4_potential}. Ratios larger than 1 (horizontal dashed line) indicate that the fits including the $\gamma$ correction fit provides a better description of the data.}
    \label{fig:App_potential_chi2lingamma}
\end{figure}

\paragraph*{Linear growth vs. growth with the $\gamma$-correction---}
We estimate the parameters $\sigma$ and $\gamma$ of the confining potential [Eq.~\eqref{eq:V(R)_gamma}] using the systematic procedure described in App.~\ref{sec:App_procedure_fit}. In Fig.~\ref{fig:App_potential_chi2lingamma}, we compare purely linear fits with the fits including the L\"uscher correction term $\gamma$ by considering the ratio between the corresponding reduced chi-squared values, $\chi^2_r=\chi^2/N_{\rm d.o.f}$ (with $N_{\rm d.o.f}=2$ and $N_{\rm d.o.f}=3$ for the two fits, respectively), for all fitted intervals $\alpha$. This ratio is always greater than $1$, indicating that including the $\gamma$ correction provides a better description of the data. 
\paragraph*{Details on the estimation of the universal L\"usher term---}
Here, we provide details on the estimation of the universal L\"usher term $\gamma_0$ shown in Fig.~\ref{fig:Fig5_gamma0} in the main text. We estimate $\gamma_0$ by constructing the distribution obtained from all possible combinations of the confining-potential parameters $\gamma$ and $\sigma$ with the parameter $A$ extracted from the logarithmic growth of the string width. Specifically, we combine estimates obtained from different fitting intervals for the potential and the width, as well as from the different sublattices $k$. For each pair of intervals $(\alpha_1, \alpha_2)$ and each sublattice $k$, we define
\begin{equation}
    \gamma_{0, (\alpha_1, \alpha_2, k)} = \frac{1}{2\pi} \frac{\gamma_{\alpha_1}}{\sigma_{\alpha_1}} \frac{1}{A_{\alpha_2, k}},
    \label{eq:app_gamma_0}
\end{equation}
where $\gamma_{\alpha_1}$ and $\sigma_{\alpha_1}$ are the parameters extracted from the fit of the confining potential over the interval $\alpha_1$, and $A_{\alpha_2, k}$ is the width-growth parameter obtained from the fit over the interval $\alpha_2$ for the sublattice $k$.

\begin{figure*}
    \centering
    \includegraphics[width=\textwidth]{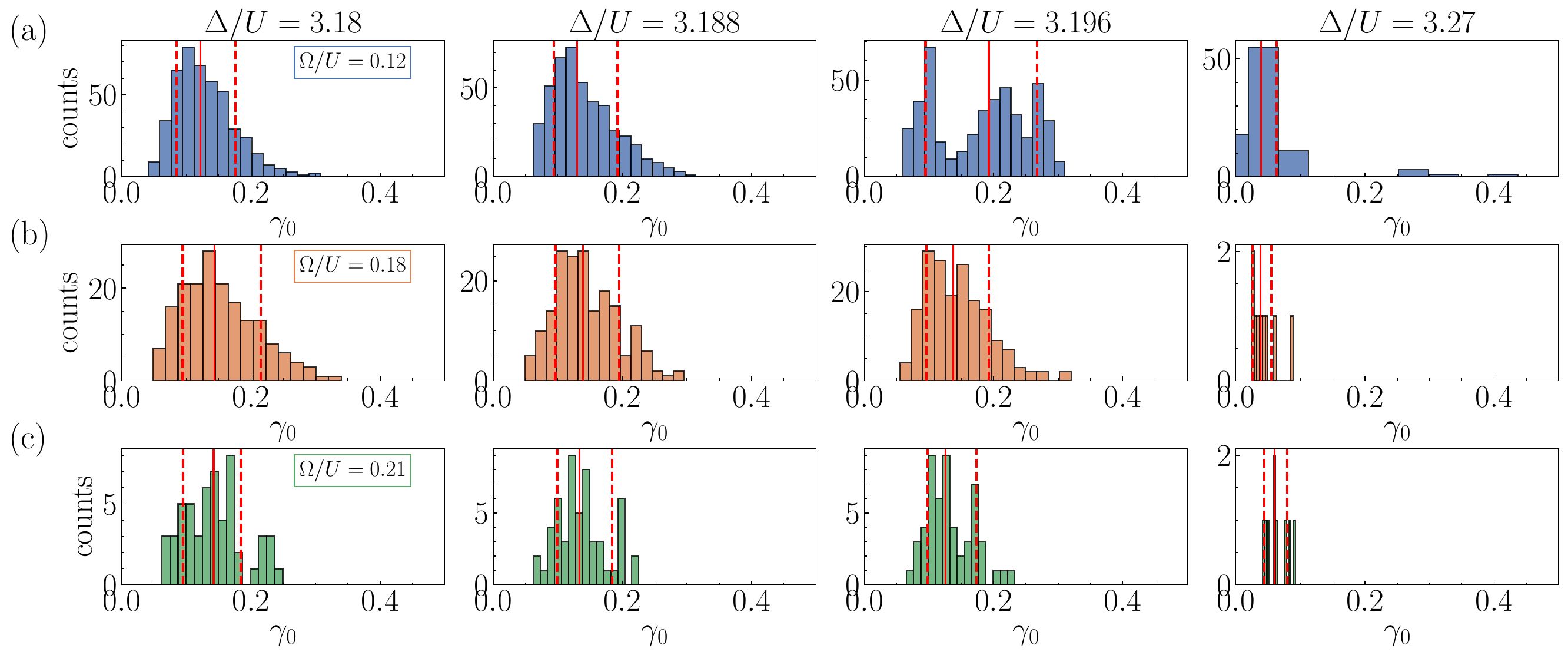}
    \caption{Distribution of the universal L\"uscher correction $\gamma_0$ for different values of $\Omega/U$ (rows) and $\Delta/U$ (columns). Each panel shows the distribution $\{\gamma_{0, (\alpha_1, \alpha_2, k)}\}_{\alpha_1, \alpha_2, k}$ obtained by combining the parameters of the confining potential fitted over the interval $\alpha_1$ with that of the string width for the sublattice $k$ fitted over the interval $\alpha_2$, according to Eq.~\eqref{eq:app_gamma_0}. The median of each distribution (solid red line) is taken as the estimate of $\gamma_0$, with the uncertainty indicated by the red dashed lines.}
    \label{fig:gamma0_histograms}
\end{figure*}

In Fig.~\ref{fig:gamma0_histograms}, we show the distributions of $\{\gamma_{0, (\alpha_1, \alpha_2, k)}\}_{\alpha_1, \alpha_2, k}$ for the different values of $\Omega/U$ (different rows) and some values of $\Delta/U$ (different columns). Each distribution is centered around its median value (solid red line), which we take as an estimate of $\gamma_0$, with uncertainty indicated by the red dashed lines. For large $\Delta/U$, the distributions become narrow and shifted towards zero, consistent with the absence of a L\"uscher correction deep in $2/3$ phase. As $\Delta/U$ is lowered, the distribution acquires a finite median value, consistent with the universal L\"uscher correction term [see Fig.~\ref{fig:Fig5_gamma0} in the main text].


\section{Including interactions up to fifth-nearest neighbors}\label{sec:App_5NN}
\begin{figure}
    \centering
    \includegraphics[width=\columnwidth]{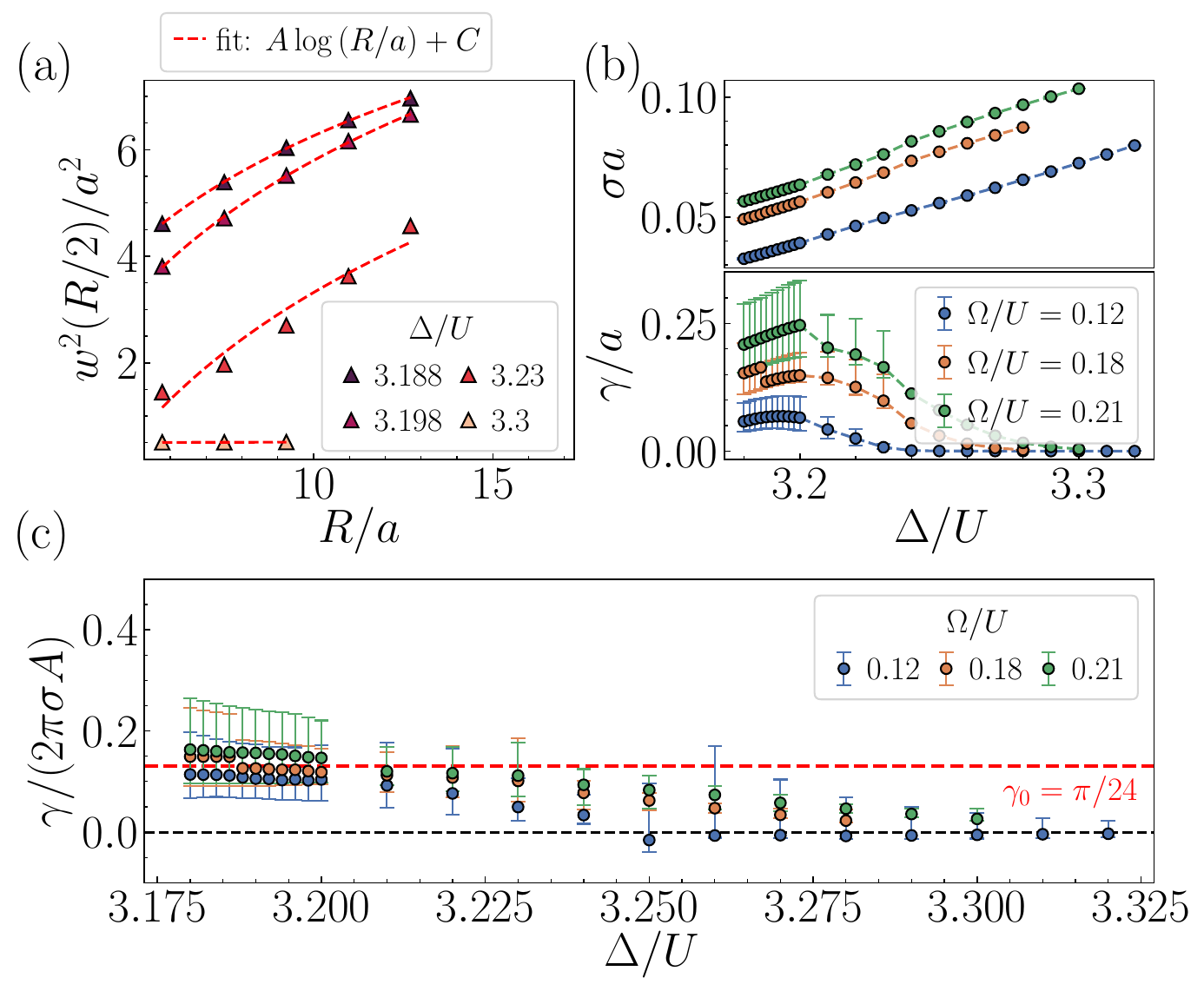}
    \caption{Analyses including interactions up to the fifth-nearest neighbors. (a) String width $w^2(R/2)$ for the $k=0$ sublattice as a function of $R$ for different $\Delta/U$. Red lines indicate fits to $A\log{R}+C$. (b) String tension $\sigma$ and $\gamma$ correction to the linear confining potential as a function of $\Delta/U$ for different $\Omega/U$. (d) Estimate of the universal L\"uscher correction $\gamma_0$ as a function of $\Delta/U$ for different $\Omega/U$. The dashed horizontal line indicate the universal value $\gamma_0=\pi/24$ for a rough string in 2D.}
    \label{fig:Fig_app_5NN}
\end{figure}

Here, we repeat the analyses presented throughout the main text, now extending the interactions up to the fifth-nearest neighbor, and using a bond dimension of $\chi=900$. We find that the main conclusions of our work remain unchanged. We extract the universal $\gamma_0$ term from  both the logarithmic growth of the string width and the correction to the linear confining potential, and find it to remain consistent with the expected value $\pi/24$.

As discussed in the Supplementary Material of Ref.~\cite{Bombieri_2025}, including interactions up to the fifth-nearest neighbor shifts the position of the DQCP shifts to a larger $\Delta/U$ value---from approximately $3.158$ to around $3.1796$. For a fixed $\Delta/U$ within the $2/3$ phase, we thus expect enhanced string fluctuations, resulting in stronger boundary effects. Consequently, we anticipate larger uncertainties arising both from these boundary effects and from the finite bond dimension, since DMRG convergence becomes more demanding in the presence of extended interactions.

In Fig.~\ref{fig:Fig_app_5NN}(a), we show the string width at half the distance between the charges for the  $k=0$ sublattice as a function of distance, for different 
$\Delta/U$ values at $\Omega/U=0.18$. As anticipated due to the shifted DQCP, the string width is larger for a fixed $\Delta/U$ compared to the case with interactions up to the third-nearest neighbor [cf.~Fig.~\ref{fig:Fig3_width_log}(a)]. Nevertheless, the behavior remains consistent with a constant for large 
$\Delta/U$ and a logarithmic growth for small $\Delta/U$ (red lines). 
In Fig.~\ref{fig:Fig_app_5NN}(b), we show the estimates of the string tension $\sigma$ and the $\gamma$ correction to the linear confining potential as a function of $\Delta/U$. As expected, $\sigma$ decreases when extending the interaction range [cf.~Fig.~\ref{fig:Fig4_potential}(b)].

Finally, in Fig.~\ref{fig:Fig_app_5NN}(d), we show that the universal correction $\gamma_0$ remains compatible with the universal L\"usher value $\gamma_0=\pi/24$ for all considered $\Omega/U$ at small $\Delta/U$. Compared to the results in the main text [Fig.~\ref{fig:Fig5_gamma0}],  uncertainties are larger, reflecting stronger boundary effects and the increased difficulty in achieving full convergence for longer-range interactions.

\section{Further details on the dynamics}\label{sec:App_dynamics}
\begin{figure}
    \centering
\includegraphics[width=\columnwidth]{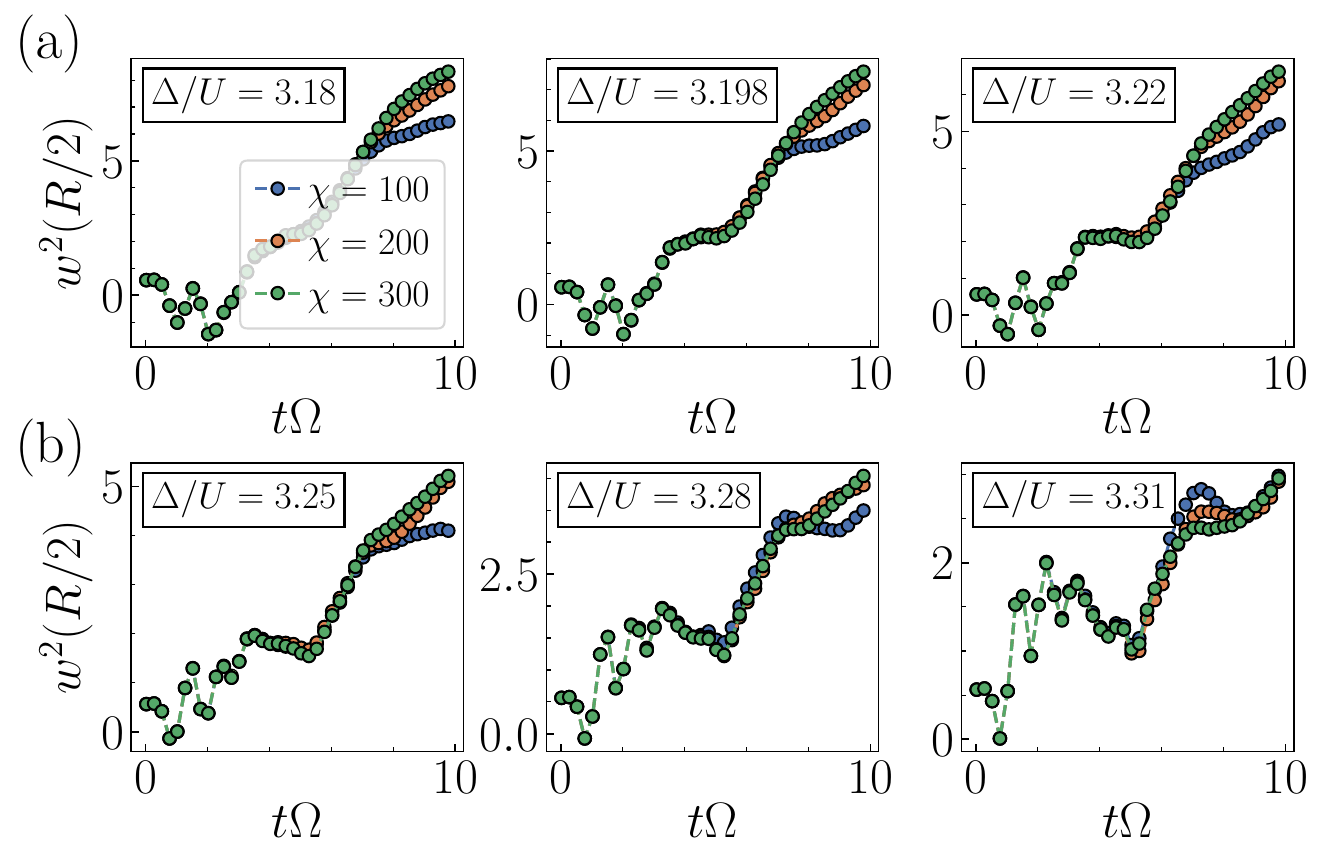}
    \caption{Convergence of the string width  $w^2(R/2)$ with respect to the bond dimension $\chi$, shown as a function of time $t\Omega$ [c.f. Fig.~\ref{fig:Fig7_dynamics_w2}(a)]. Different panels correspond to quenches to different values of $\Delta/U$.}
    \label{fig:App_dynamics_w2_chi}
\end{figure}

In Fig.~\ref{fig:App_dynamics_w2_chi}, we provide additional data on the convergence of the string width $w^2(R/2)$ shown in Fig.~\ref{fig:Fig7_dynamics_w2}(a) with respect to the bond dimension $\chi$, for quenches to $\Omega/U=0.18$ and different values of $\Delta/U$.

\bibliography{bibliography}

\end{document}